\documentclass[letterpaper, 10 pt, conference]{ieeeconf}  

\IEEEoverridecommandlockouts                              

\usepackage{amssymb,amsmath,amsfonts,amsthm}
\usepackage{algorithmic}
\usepackage{bm}
\usepackage{color}
\usepackage{graphicx}
\usepackage{cite}
\usepackage{booktabs}
\usepackage{multirow}
\usepackage{threeparttable}
\usepackage{enumerate}
\usepackage[english]{babel}
\usepackage[ruled, lined, linesnumbered, commentsnumbered, longend]{algorithm2e}
\usepackage{nomencl}
\usepackage{multicol}
\usepackage{commath}
\usepackage{makecell}
\usepackage{blindtext}
\usepackage{soul}
\usepackage{comment}
\usepackage{tabularx}
\usepackage[dvipsnames]{xcolor}
\usepackage{colortbl}
\usepackage{url}
\usepackage{csquotes}
\usepackage[export]{adjustbox}
\usepackage{physics}
\usepackage{mathtools}
\usepackage{balance}

\makeatletter
\let\NAT@parse\undefined
\makeatother
\usepackage[hidelinks]{hyperref}

\theoremstyle{definition}
\newtheorem*{thm*}{Theorem}
\newtheorem{thm}{Theorem}
\newtheorem{defn}{Definition}
\newtheorem{rem}{Remark}
\newtheorem{prop}{Proposition}
\newtheorem{lem}{Lemma}
\newtheorem{assum}{Assumption}

\newtheorem{example}{Example}
\newtheorem*{notation}{Notation}

\newcommand{\vect}[1]{\mathbf{#1}}
\newcommand{\matr}[1]{\mathbf{#1}}
\newcommand{\tran}{\mathsf{T}}

\title{\LARGE \bf
On the Equivalence of Koopman Eigenfunctions and Commuting Symmetries
}

\author{Xinyuan Jiang$^1$ and Yan Li$^2$
\thanks{This work was partly supported by the Office of Naval Research under Award N00014-22-1-2504.} 
\thanks{$^1$Xinyuan Jiang is an independent researcher {\tt\small j\_jxy@outlook.com}}%
\thanks{$^2$Yan Li is with the School of EECS, The Pennsylvania State University, University Park, PA 16802, USA {\tt\small yql5925@psu.edu}}%
}

\begin{document}

\maketitle
\thispagestyle{empty}
\pagestyle{empty}

\begin{abstract}
The Koopman operator framework offers a way to represent a nonlinear system as a linear one. The key to this simplification lies in the identification of eigenfunctions. While various data-driven algorithms have been developed for this problem, a theoretical characterization of Koopman eigenfunctions from geometric properties of the flow is still missing. This paper provides such a characterization by establishing an equivalence between a set of Koopman eigenfunctions and a set of commuting symmetries---both assumed to span the tangent spaces at every point on a simply connected open set. Based on this equivalence, we derive an explicit formula for the principal Koopman eigenfunctions and prove its uniform convergence on the region of attraction of a locally asymptotically stable equilibrium point, thereby offering a constructive method for computing Koopman eigenfunctions.


\end{abstract}

\section{Introduction}

Nonlinear dynamical systems are often studied through the evolution of functions of the state, i.e., observables. The Koopman operator framework offers a linear perspective on nonlinear dynamics. It describes the evolution of observables via an infinite-dimensional linear operator. Koopman eigenfunctions, in particular, serve as powerful tools for spectral analysis, dimensionality reduction, and model discovery~\cite{mezic2021koopman,brunton2021modern}.

Despite their importance, the computation of Koopman eigenfunctions remains challenging. Data-driven methods, such as extended dynamic mode decomposition (EDMD) and related variants~\cite{williams2015data,korda2020optimal,colbrook2023residual}, provide practical approximations from simulation or experimental data. Recent contributions have introduced finite-data error bounds, which quantify the reliability of these approximations from both a learning-theoretic perspective~\cite{nuske2023finite} and control-oriented contexts~\cite{mamakoukas2021derivative}. These results underscore the potential of Koopman-based methods. At the same time, they highlight a fundamental limitation arising from the absence of analytical characterizations of eigenfunctions.

One emerging direction to address this limitation is to incorporate system symmetries into Koopman learning. By enforcing equivariance or symmetry constraints, researchers have extended Koopman-based models to dynamical systems with geometric or physical structure~\cite{sinha2020koopman,peitz2023equivariance,sharma2016correspondence,marensi2023symmetry,weissenbacher2022koopman,ordonez2024dynamics,mesbahi2021nonlinear,salova2019koopman,jiang2024modularized}. These approaches demonstrate that symmetry constraints can improve the stability of data-driven algorithms and enhance their interpretability. Nevertheless, most existing methods impose symmetry constraints algorithmically, without establishing a principled connection between Koopman eigenfunctions and intrinsic system symmetries.

In parallel, geometric control theory provides a rigorous framework for analyzing nonlinear dynamical systems through differential geometry. Fundamental concepts such as Lie group, Lie algebra, and foliation underlie classical results in observability, controllability, and reachability~\cite{agrachev2013control}. These tools naturally capture the role of symmetries in establishing global linearization through state immersion~\cite{menini2009linearization}. However, their relationship to the Koopman operator framework has not been systematically established.

This paper establishes a connection between the two perspectives by giving a direct equivalence between Koopman eigenfunctions and commuting symmetries. Specifically, we show that on a simply connected open domain, the gradients of Koopman eigenfunctions are mapped bijectively to commuting symmetry generators that span the tangent spaces. This equivalence provides a structural characterization of Koopman eigenfunctions that complements existing data-driven methods. Based on it, we also derive an explicit formula for Koopman eigenfunctions. We then prove that it converges uniformly on compact subsets of a region of attraction. In summary, the main contributions of this paper are:
\begin{itemize}
    \item We establish the equivalence between Koopman eigenfunctions and commuting symmetries by providing a bijective mapping between them.
    \item We derive an explicit formula and prove its uniform convergence to Koopman eigenfunctions on compact subsets of a region of attraction.
\end{itemize}

This paper is organized as follows. Section~\ref{sec_symmetry} covers geometric preliminaries. Section~\ref{sec_Koopman} connects Koopman eigenfunctions to commuting symmetries. Section~\ref{sec_formulas} specializes to the region of attraction and derives the computable limit (\ref{E:s6-9}). Section~\ref{sec_conclusion} offers concluding remarks.

\begin{notation}
The imaginary unit is $j$. A function is denoted by lower-case italic (e.g., $\psi$). A vector field~\cite{lee2003smooth} is denoted by upper-case italic (e.g., $F$). The exceptions are $U$, $A$, and $K$, which are subsets of $\mathbb{R}^n$.
The zero and one vectors in $\mathbb{R}^n$ are $0_n$ and $1_n$, respectively.
The standard basis of $\mathbb{R}^n$ is $(\vect e_1,\ldots,\, \vect e_n)$. $(\cdot)^*$ denotes conjugate transpose, and $(\cdot)^\tran$ denotes transpose. For a complex-valued function $g: \mathbb{R}^n \to \mathbb{C}$, the gradient $\nabla g(\vect x)$ is a column vector such that
\begin{equation*}
    \frac{\partial g(\vect x)}{\partial \vect x} \vect v = \nabla g(\vect x)^* \vect v,\quad \forall \vect v \in \mathbb{R}^n.
\end{equation*}
For a vector field $X(\vect x)$ on $\mathbb{R}^n$, the Jacobian $\nabla X(\vect x)$ is a square matrix such that \begin{equation*}
    \frac{\partial X(\vect x)}{\partial \vect x} \vect v = \nabla X(\vect x) \vect v,\quad \forall \vect v \in \mathbb{R}^n.
\end{equation*}
\end{notation}

\section{Preliminaries} \label{sec_symmetry}


For a vector field $F$ on $\mathbb{R}^n$, there is an associated local flow $\Phi_F(t, \vect x)$ such that $\frac{\partial}{\partial t} \Phi_F(t, \vect x) = F(\Phi_F(t, \vect x))$ for all $\vect x \in \mathbb{R}^n$, $t \in \mathrm I_{\vect x} \subset \mathbb{R} \ni 0$. In particular, $\vect x(t) = \Phi_F(t, \vect x_0)$ is the trajectory of the ODE system,
\begin{equation} \label{E:system}
    \dot{\vect x} = F(\vect x),\quad \vect x \in \mathbb{R}^n,
\end{equation}
starting from the initial condition $\vect x_0 \in \mathbb{R}^n$.
If $F(\vect x)$ is $\mathcal C^\infty$-smooth, then the trajectory from every initial condition is unique. The flow $\Phi_F(t, \vect x)$ is said to be complete if, for each $\vect x\in \mathbb{R}^n$, $\Phi_F(t, \vect x)$ belongs to $\mathbb{R}^n$ for all $t\in \mathbb{R}$; that is, it is invariant in $\mathbb{R}^n$.
For simplicity, we assume that the flow $\Phi_F(t, \vect x)$ is complete; this holds for many physically relevant models~\cite{angeli1999forward}. We will also use the notation $\Phi_F^t(\vect x)$ for $\Phi_F(t, \vect x)$.

A coordinate-independent way to describe the vector field $F(\vect x)$ is provided by the Lie derivative $\mathcal L_F: \mathcal C^\infty(\mathbb{R}^n; \mathbb{C}) \to \mathcal C^\infty(\mathbb{R}^n; \mathbb{C})$ defined as
\begin{equation} \label{E:lie}
    \mathcal L_F \varphi(\vect x) = \nabla \varphi(\vect x)^* F(\vect x).
\end{equation}
The linear operator $\mathcal L_F$ describes the infinitesimal change of a function $\varphi(\vect x)$ along the flow $\Phi_F(t, \vect x)$, independently of coordinate choice.\footnote{The coordinates for (\ref{E:system}) are the state vector $\vect x$, according to which the partial derivatives and the gradient are defined.} 
Later we will see that $\mathcal L_F$ generates the Koopman operator.

In the remainder of this section, we recall some preliminary results on commuting and conservative vector fields.

\subsection{Definition of Trajectory Symmetries}

For the system (\ref{E:system}), each symmetry is a diffeomorphism of $\mathbb{R}^n$ that leaves the flow of (\ref{E:system}) invariant. 
Symmetries are usually studied from the group perspective: Since composing two symmetries yields another symmetry, the set of all symmetries forms a group under composition.
Of particular interest are (sub)sets of symmetries that form one-parameter groups, i.e., homomorphic groups to the additive group $\mathbb{R}$.

\begin{defn} \label{def_one_para}
A one-parameter group is a mapping $\mathcal T(s)$ from the additive group $\mathbb{R}$ to a Lie group $\mathbb{G}$ that is a group homomorphism, i.e., $\mathcal T(r) \mathcal T(s) = \mathcal T(r+s)$. The infinitesimal generator of $\mathcal T(s)$ is $G = \frac{\partial \mathcal T}{\partial s}(0)$.
\end{defn}

For any one-parameter group $\mathcal T(s)$, it holds that, for each $s'\in \mathbb{R}$,
\begin{equation} \label{E:induction}
    \frac{\partial\mathcal T}{\partial s}(s') = \frac{\partial \mathcal T}{\partial s}(0) \mathcal T(s') = G \mathcal T(s'),
\end{equation}
where the group property is used in the first equality. Noting that $\mathcal T(0) = \mathrm{Id}$ and (\ref{E:induction}), we obtain $\mathcal T(s) = \mathrm{exp}(sG)$~\cite{lee2003smooth}.
The primary example we have in mind is the complete flow $\Phi_F^t$ of (\ref{E:system}), which is a one-parameter group of diffeomorphisms of $\mathbb{R}^n$ generated by $F(\vect x)$. 
Formally, symmetries are defined using local one-parameter groups. These are groups defined for $s$ in a neighborhood of $0$, which accounts for the fact that the flow of a vector field may be incomplete.

\begin{defn} \label{def_sym}
On an open subset $U\subset \mathbb{R}^n$, the local flow $\Phi_G^s$ generated by a vector field $G$ is said to be a symmetry of (\ref{E:system}) if $\Phi_G^s \Phi^t_F \vect x = \Phi^t_F \Phi_G^s \vect x$ for all $\vect x \in U$, $t \in \mathbb{R}$, and $s \in \mathrm I_{\vect x}$.
\end{defn}

Equivalently, the commutativity relationship between $\Phi_G^s$ and $\Phi_F^t$ means that $\Phi_G^s$ is a local one-parameter group of diffeomorphisms that map each trajectory of (\ref{E:system}) to either the same or another trajectory of (\ref{E:system}). As a simple example, rotational symmetry on $\mathbb{R}^2$ is generated by the vector field $G(\vect x) = \big[0\; {-}1;1\; 0\big] \vect x$.

\begin{lem} \label{prop_equiv_def}
Consider an open subset $U \subset \mathbb{R}^n$ and a vector field $G(\vect x)$ on $U$.
The associated local one-parameter group of diffeomorphisms $\Phi_G^s: U\to U$ is a symmetry if and only if the following equivalent conditions hold:
\begin{align}
    &\nabla \Phi_G^s(\vect x) F(\vect x) = F(\Phi_G^s(\vect x)),\quad \forall \vect x \in U,\, s\in \mathrm I_{\vect x}; \label{E:s1-1}
\end{align}
and
\begin{align}
    &[G, F](\vect x) = \nabla G(\vect x) F(\vect x) - \nabla F(\vect x) G(\vect x) = 0_n,\quad \forall \vect x \in U. \label{E:s1-2}
\end{align}
\end{lem}

The proof is omitted; see, e.g.,~\cite{lee2003smooth}.

\begin{rem} \label{rem_flow_box}
By the flow-box theorem~\cite{isidori1985nonlinear}, near each non-equilibrium point $\vect x'$ of (\ref{E:system}), there is a local diffeomorphism $\widehat{\vect z} = Z^{-1}(\vect x)$ such that, in $\widehat{\vect z}$-coordinates, the flow is the constant flow $\dot{\widehat{\vect z}} = \vect e_1$.
Then, a set of commuting symmetry generators can be taken as $\vect e_i$ in $\widehat{\vect z}$-coordinates, which correspond to $\nabla Z(\widehat{\vect z}) \vect e_i$ in $\vect x$-coordinates. However, identifying global symmetries for a general nonlinear system remains challenging~\cite{kooij1993algebraic,christopher1994invariant,walcher2000plane,goriely2001integrability}.
\end{rem}

It is important to note that in Definition~\ref{def_sym} the flow $\Phi_G^t$ of the symmetry is not necessarily complete even if the flow of (\ref{E:system}) is. This makes the flow $\Phi_G^t$ itself quite cumbersome to use, and is probably a reason why it is customary to call a vector field $G$ satisfying the commuting condition (\ref{E:s1-2}) a symmetry, instead of its local flow.

To accommodate the complex Koopman spectrum later, we slightly generalize the definition of symmetry.
A (generalized) vector field $G: U\to \mathbb{C}^n$ is said to be a (complex) symmetry for (\ref{E:system}) if both its real and imaginary parts are vector fields that are symmetries. It is defined by the same condition (\ref{E:s1-2}) due to the bilinearity of the Lie bracket.

\subsection{Commuting and Conservative Local Frames}

Consider a set of $n$ (generalized) vector fields defined on $U\subset \mathbb{R}^n$,
\begin{equation} \label{E:vector_fun}
    \big[ E_1(\vect x) \; \cdots \; E_n(\vect x) \big] \in \mathbb{C}^{n\times n}
\end{equation}
that form a full-rank, commuting set, i.e., $[E_i, E_k](\vect x) = 0_n$. The goal is to show that the $n$ commuting vector fields can be mapped bijectively to the gradients of $n$ scalar functions. Conceptually, the trajectory lines of the commuting vector fields define a set of curvilinear coordinates. The coordinate values at each point are exactly the values of the associated scalar functions.

To make the relationship between commutativity and integrability precise, we introduce some additional definitions.
A vector field $X$ on $U$ is said to be conservative if there is a scalar function $m: U \to \mathbb{C}$ such that $\nabla m(\vect x) = X(\vect x)$ for all $\vect x \in U$. This property of the vector field can be checked by applying the Poincar\'e lemma~\cite{lee2003smooth}.

\begin{lem} \label{lem_poin}
A vector field $X: U \to \mathbb{C}^n$ defined on a simply connected open subset $U$ of $\mathbb{R}^n$ is conservative if and only if 
\begin{equation} \label{E:integrability}
    \nabla X(\vect x) = \nabla X(\vect x)^\tran,\quad \forall \vect x \in U.
\end{equation}
\end{lem}

\begin{defn}
A set of $n$ (generalized) vector fields (\ref{E:vector_fun}) on $U \subset \mathbb{R}^n$ is called a local frame for $U$ if the square matrix (\ref{E:vector_fun}) is full-rank for each $\vect x \in U$.
\end{defn}

The complete theorem is stated as follows.

\begin{thm} \label{prop_equiv}
On a simply connected open subset $U \subset \mathbb{R}^n$, a commuting local frame $\big[ E_1(\vect x) \; \cdots\; E_n(\vect x) \big] \in \mathbb{C}^{n\times n}$ is mapped to a conservative local frame $\big[ X_1(\vect x) \; \cdots\; X_n(\vect x) \big] \in \mathbb{C}^{n\times n}$ by 
\begin{align} \label{E:inverse}
    \big[X_1(\vect x)\; \cdots\; X_n(\vect x) \big]^* = \big[E_1(\vect x)\; \cdots\; E_n(\vect x) \big]^{-1},
\end{align}
and vice versa.
\end{thm}

The proof is given in Appendix~\ref{proof_prop_equiv}.


\section{Koopman Eigenfunctions From Commuting Symmetries} \label{sec_Koopman}

Koopman eigenfunctions for (\ref{E:system}) are functions that evolve exponentially along the flow, much like eigenvectors for a linear system. As a result, under a full-rank assumption, they can serve as coordinates that linearize the nonlinear dynamics on a nonlocal set, capturing nonlocal behavior. Hence, identifying the eigenfunctions has become a central objective in modern nonlinear systems research.

We will show that any set of linearly independent (in the sense of a local frame) Koopman eigenfunctions has an associated commuting local frame of symmetries.
Specifically, we will sharpen (\ref{E:inverse}) to prove that $\big[X_1(\vect x)\;\cdots\; X_n(\vect x) \big]$ in the LHS is a local frame of gradients of logarithms of Koopman eigenfunctions for (\ref{E:system}) if and only if $\big[E_1(\vect x)\; \cdots\; E_n(\vect x) \big]$ in the RHS is a commuting local symmetry frame for (\ref{E:system}).

\subsection{Koopman Eigenfunctions} \label{sec_1}

Assuming that the flow of (\ref{E:system}) is complete, the associated Koopman operator is a one-parameter group of linear operators $\mathcal K(t): \mathcal C^\infty(U; \mathbb{C}) \to \mathcal C^\infty(U; \mathbb{C})$ such that~\cite{mezic2020spectrum}
\begin{equation*}
    \mathcal K(t) \varphi(\vect x) = [\varphi \circ \Phi_F] (t, \vect x) = \varphi( \Phi_F(t, \vect x)),\quad t \in \mathbb{R}.
\end{equation*}
The infinitesimal generator is
\begin{align*}
    \frac{\partial \mathcal K}{\partial t}(0) \varphi(\vect x) = \frac{\partial \varphi\circ \Phi_F}{\partial t}(0, \vect x) &= \nabla \varphi(\vect x)^* F(\vect x) = \mathcal L_F \varphi(\vect x),
\end{align*}
which is the Lie derivative.

\begin{defn}[Definition~5.1 of~\cite{mezic2020spectrum}] \label{def_eigen}
A function $\psi: U\to \mathbb{C}$ is said to be an open Koopman eigenfunction for (\ref{E:system}) on the open subset $U\subset \mathbb{R}^n$ if, for some $\mu \in \mathbb{C}$ (the eigenvalue), it holds that
\begin{equation} \label{E:eigen}
    \mathcal L_F \psi(\vect x) = \nabla \psi(\vect x)^* F(\vect x) = \mu \psi(\vect x),\quad \forall \vect x \in U.
\end{equation}
\end{defn}

\begin{lem} \label{lem_prop}
An eigenpair $(\mu, \psi(\vect x)) \in \mathbb{C} \times \mathcal C^\infty(U; \mathbb{C})$ of the Koopman generator on some open subset $U \subset \mathbb{R}^n$ has the following properties:
\begin{enumerate}
    \item[i)] The eigenfunction evolves exponentially along the flow:
    $\mathcal K(t) \psi(\vect x) = \psi(\Phi_F(t, \vect x)) = e^{\mu t} \psi(\vect x),\, \forall \vect x \in U$.
    \item[ii)] Given two eigenpairs $(\mu_1, \psi_1(\vect x)), (\mu_2, \psi_2(\vect x))$, there is an eigenpair:
    $(\mu_1 + \mu_2, \psi_1(\vect x) \psi_2(\vect x))$.
    \item[iii)] Given an eigenpair $(\mu, \psi(\vect x))$ such that $\psi(\vect x) \neq 0$, there is the following set of eigenpairs:
    $\big\{(k \mu, \psi(\vect x)^k )\mid k \in \mathbb{N} \big\}$.
\end{enumerate}
\end{lem}

The proof is omitted. 

The feasibility of Koopman eigenfunctions in representing the nonlinear dynamics (\ref{E:system}) can be understood in two ways. First,
given $n$ Koopman eigenfunctions $\psi_i(\vect x)$ with the associated eigenvalues $\mu_i$, we can write
\begin{equation} \label{E:recov}
    \big[ \nabla \psi_1(\vect x) \;\cdots\; \nabla \psi_n(\vect x) \big]^* F(\vect x) = \begin{bmatrix}\mu_1 \psi_1(\vect x) \\
    \vdots \\ \mu_n \psi_n(\vect x) \end{bmatrix}.
\end{equation}
If $\big[ \nabla \psi_1(\vect x) \;\cdots\; \nabla \psi_n(\vect x) \big]$ is full-rank for all $\vect x \in U$, then we can solve for $F(\vect x)$ from (\ref{E:recov}) through matrix inverse. Second, in the eigenfunction coordinates, the dynamics can be expressed as the linear system $\dot{\vect z} = \matr A \vect z$, where $\vect z = \big[ \psi_1(\vect x) \;\cdots\; \psi_n(\vect x) \big]^\tran$ and $\matr A = \mathrm{diag}(\mu_1,\ldots,\, \mu_n)$. The second aspect potentially enables the application of linear control to nonlinear systems.

\subsection{Main Result} \label{sec_disc}

From properties ii) and iii) in Lemma~\ref{lem_prop}, Koopman eigenfunctions combine with each other through multiplication, which makes it difficult to define linear independence for a set of $n$ Koopman eigenfunctions. We simplify the problem by considering the logarithms of the Koopman eigenfunctions. More precisely, assuming that the Koopman eigenfunctions are nonzero on $U$, we will characterize the gradients of their logarithms, which are referred to as conservative linearizing vector fields.

\begin{defn}
On a simply connected open subset $U \subset \mathbb{R}^n$,
a (generalized) vector field $X: U \to \mathbb{C}^n$ is said to be a conservative linearizing vector field if it is conservative and
\begin{equation} \label{E:flat_prop}
    \nabla X(\vect x)^* F(\vect x) + \nabla F(\vect x)^* X(\vect x) = 0_n,\quad \forall \vect x \in U.
\end{equation}
\end{defn}

\begin{rem}
We use the term “conservative linearizing vector field” because these vector fields correspond to coordinates under which the dynamics reduce to constant translations.
\end{rem}

Let $X(\vect x) = \nabla m(\vect x)$ for $\vect x \in U$. Eq. (\ref{E:flat_prop}) is, by definition, the gradient of the following equation,
\begin{equation} \label{E:flat}
    \nabla m(\vect x)^* F(\vect x) = c,\quad \forall \vect x \in U.
\end{equation}
To connect $m(\vect x)$ to a Koopman eigenfunction, let us consider the function $\psi(\vect x) = \exp(m(\vect x))$, i.e., the exponential of $m(\vect x)$. Then, we have
\begin{equation} \label{E:KEF_flat}
    \nabla \psi(\vect x) = \psi(\vect x)^* \nabla m(\vect x),
\end{equation}
and it is readily seen that $\psi(\vect x)$ is a Koopman eigenfunction with the eigenvalue given by the constant $c$ in (\ref{E:flat}). The main result on the equivalence of Koopman eigenfunctions and commuting symmetries is stated below.

\begin{prop} \label{prop_main1}
On a simply connected open subset $U\subset \mathbb{R}^n$, a local frame of conservative linearizing vector fields $\big[X_1(\vect x)\; \cdots\; X_n(\vect x)\big] \in \mathbb{C}^{n\times n}$ is mapped to a commuting local frame of symmetries $\big[E_1(\vect x)\; \cdots\; E_n(\vect x)\big] \in \mathbb{C}^{n\times n}$ by (\ref{E:inverse}), and vice versa.
\end{prop}

The proof is given in Appendix~\ref{proof_prop_main1}.

The relationship between Koopman eigenfunctions and commuting symmetries can now be summarized as follows:
\begin{gather*}
\text{A linearly independent set of $n$ Koopman eigenfunctions} \\ \phantom{\; \text{\small via (\ref{E:KEF_flat}) assuming $\psi_i(\vect x) \neq 0$}} \mathbin{\rotatebox[origin=c]{90}{$\Longleftarrow$}}\; \text{\small via (\ref{E:KEF_flat}) assuming $\psi_i(\vect x) \neq 0$} \\
\text{A local frame of conservative linearizing vector fields} \\
\phantom{\; \text{\small Proposition~\ref{prop_main1}}} \mathbin{\rotatebox[origin=c]{90}{$\Longleftrightarrow$}}\; \text{\small Proposition~\ref{prop_main1}} \\
\text{A local frame of commuting symmetries}
\end{gather*}
All are assumed to be defined on a simply connected open subset $U\subset \mathbb{R}^n$. 
In addition, the Koopman eigenfunctions can be computed by integrating their gradients, which is well-defined on a simply connected subset $U \subset \mathbb{R}^n$ by the Poincar\'e lemma. By choosing $U$ to be simply connected, we have avoided the problem that the complex logarithm is multi-valued around a singularity, i.e., points where the Koopman eigenfunction is zero. The singularities represent limit sets of the system.
If these singularities are allowed to puncture $U$, then on the punctured domain, the complex logarithms of Koopman eigenfunctions can be multi-valued around each singularity, whereas the Koopman eigenfunctions usually remain single-valued~\cite{kvalheim2025global}.

\begin{rem}[Comparison to flow-box theorem]
Recall from Remark~\ref{rem_flow_box} that, if there is a commuting local frame of symmetries $\big[E_1(\vect x)\;\cdots\; E_n(\vect x)\big]$ for (\ref{E:system}), then there exist inverse coordinates $\widehat{\vect z} = Z^{-1}(\vect x)$ whose gradients $\nabla \widehat{\vect z} = \nabla Z^{-1}(\vect x)$ coincide with the local frame of commuting symmetries. In comparison, Proposition~\ref{prop_main1} states that $\nabla \vect z = \nabla Z(\vect x)$ are conservative linearizing vector fields for (\ref{E:system}). The latter correspond to the gradients of logarithms of Koopman eigenfunctions, and offer nonlocal representations of (\ref{E:system}) as a linear system (at least valid on the simply connected open set $U$).
\end{rem}


\begin{rem}
To gain some further insights into the Koopman eigenfunctions, we replace the last symmetry generator in the RHS of (\ref{E:inverse}) by the system vector field; that is,
\begin{align} \label{E:sym}
    \big[E_1(\vect x)\; \cdots\;  E_{n-1}(\vect x)\; F(\vect x)\big].
\end{align}
Plugging (\ref{E:sym}) into (\ref{E:inverse}), we obtain
that the first $n-1$ logarithms of Koopman eigenfunctions have $c_i = 0$, while the last logarithm of Koopman eigenfunction has $c_n = 1$. Two comments are in order:
\begin{enumerate}
    \item Firstly, by (\ref{E:sym}) and (\ref{E:inverse}), $n-1$ commuting symmetry generators that are linearly independent from $F(\vect x)$ are associated with $n-1$ logarithms of Koopman eigenfunctions with $c_i = 0$ and one logarithm of Koopman eigenfunction with $c_i = 1$. In this case, the first  $n-1$ are first integrals.
    \item Secondly, the trajectory of the system from an initial condition $\vect x_0 \in U$ can be constructed as follows. The orbit through $\vect x_0$ can be constructed from the $n-1$ first integrals as
    \begin{equation*}
        \Phi_F(\mathbb{R}, \vect x_0) = \bigcap_{i=1}^{n-1} \big\{\vect x\in \mathbb{R}^n \mid m_i(\vect x) = m_i(\vect x_0)\big\}.
    \end{equation*} 
    In addition, the time information of each orbit is provided by the last logarithm of Koopman eigenfunction.
\end{enumerate}
\end{rem}

Though the flow map is relatively straightforward to construct from $n-1$ commuting symmetries as we have shown above, it is difficult to compute the commuting symmetries from the flow map. This problem involves finding a partial order for the subset $U \subset \mathbb{R}^n$ by the flow map and becomes much simpler if $U$ is the region of attraction of a locally asymptotically stable equilibrium point, as we show next.

\section{Formula for Koopman Eigenfunctions on a Region of Attraction} \label{sec_formulas}

We derive an explicit formula for the principal Koopman eigenfunctions on a region of attraction by exploiting their connection to symmetries.
The non-rigorous formal intuition is given below by expressing the commuting symmetry generators on the region of attraction of a locally asymptotically stable equilibrium point based on the system's flow map through pushforward of the asymptotic symmetries near the equilibrium point. The full proof of the result is given in the Appendix.

Let $\vect x_0 = 0_n$ be a locally asymptotically stable equilibrium point for (\ref{E:system}). Denote the region of attraction by $A \subset \mathbb{R}^n$. The linearized system can be written as
\begin{align} 
    &\nabla F(\vect x_0) = \matr W^{-1} \matr\Lambda \matr W, \notag \\
    &\matr \Lambda = \mathrm{diag}(\lambda_1,\ldots,\, \lambda_n),\, \matr W = \big[\vect w_1\; \cdots\; \vect w_n \big]^*. \label{E:linearization}
\end{align}
We introduce the following assumption to ensure the existence of smooth Koopman eigenfunctions around $\vect x_0$.

\begin{assum} \label{assum}
The linearized system $\nabla F(\vect x_0)$ is Hurwitz, diagonalizable, and non-resonant, i.e., for every multi-index $\alpha \in \mathbb{N}^n$ with $\sum_{i} \alpha_i \geq 2$ and every $k$, $\sum_{i} \alpha_i \lambda_i \neq \lambda_k$, and 
$F(\vect x)$ is real-analytic in a neighborhood of $\vect x_0$.
\end{assum}

The goal is to exploit the regular asymptotic behavior of the flow near the equilibrium point.
To this end, we change to the following coordinates,
\begin{equation} \label{E:z_system}
    \vect z = Z(\vect x) = \begin{bmatrix}
        z_1(\vect x) \\
        \vdots \\
        z_n(\vect x)
    \end{bmatrix} = \begin{bmatrix}
        \frac{1}{\lambda_1} \log(\vect w_1^* \vect x) \\
        \vdots \\
        \frac{1}{\lambda_n} \log(\vect w_n^* \vect x)
    \end{bmatrix} \in \mathcal D,
\end{equation}
where 
$\mathcal D = \left\{\vect z\in \mathbb{C}^n \mid z_i \in \frac{1}{\lambda_i} \log(\vect w_i^* \vect x),\, \vect x \in \mathbb{R}^n \right\}$.
Denote by $\widetilde \Phi_F(t, \vect z)$ the flow in $\vect z$-coordinates.
As $\vect z \to \infty_n$, or equivalently as $t \to \infty$ along each trajectory, the dynamics asymptotically approach $\dot{\vect z} = 1_n$, such that the commuting symmetries approach simple translations.
In order to extend these symmetries to the entire region of attraction, recall the following lemma.

\begin{lem} \label{lem_grad_flow}
Consider the flow $\widetilde \Phi_F(t, \vect z)$, and, given $\vect z' \in \mathcal D$ and $\vect v \in T_{\vect z'} \mathcal D$, define the vector field $\widetilde E(\vect z)$ at every point in $\widetilde \Phi_F(\mathbb{R}, \vect z')$ such that
\begin{equation*}
    \widetilde E(\widetilde \Phi_F (t, \vect z')) = \nabla \widetilde \Phi_F (t, \vect z') \vect v.
\end{equation*}
Then, it holds that
\begin{equation*}
    \nabla \widetilde \Phi_F(t, \vect z) \widetilde E(\vect z) = \widetilde E(\widetilde \Phi_F(t, \vect z)),
\end{equation*}
for all $\vect z \in \widetilde \Phi_F(\mathbb{R}, \vect z')$.
\end{lem}

The proof is omitted.
In words, the pushforward $\widetilde E(\vect z)$ of a tangent vector $\vect v$ at $\vect z'$ along the trajectory of the system $\widetilde \Phi_F(t, \vect z')$, defines a symmetry generator on $\widetilde \Phi_F(\mathbb{R}, \vect z')$.

Note that the flow at the limit $\vect z \to \infty_n$ asymptotically approaches the constant flow $\dot{\vect z} = 1_n$, for which a set of commuting symmetries is given by the simple translations generated by $\big[\vect e_1\;\cdots\; \vect e_n\big]$. Their pushforwards are
\begin{align}
    \big[\widetilde E_1(\vect z)\; \cdots\; \widetilde E_n(\vect z) \big] &= \lim_{t\to\infty} \nabla \widetilde \Phi_{F}(-t, \widetilde \Phi_{F}(t, \vect z)) \big[\vect e_1\;\cdots\; \vect e_n\big] \notag \\
    &= \lim_{t\to\infty} \nabla \widetilde \Phi_{F}(-t, \widetilde \Phi_{F}(t, \vect z)) \notag \\
    &= \lim_{t\to\infty} \nabla \widetilde \Phi_{F}(t, \vect z)^{-1}. \label{E:s6-13}
\end{align} 
By Lemmas~\ref{lem_grad_flow} and \ref{prop_equiv_def}, each $\widetilde E_i(\vect z)$ commutes with the flow $\widetilde \Phi_F(t, \vect z)$. 
Moreover, $\widetilde E_i(\vect z), \widetilde E_k(\vect z)$ commute with each other because they are pushforwards of the commuting $\vect e_i, \vect e_k$ by the diffeomorphism $\lim_{t\to\infty} \widetilde \Phi_{F}(-t, \widetilde \Phi_F(t, \vect z))$.
Substituting (\ref{E:s6-13}) into (\ref{E:inverse}), we obtain a set of conservative linearizing vector fields as
\begin{align}
    &\big[ \widetilde X_1(\vect z)\;\cdots\; \widetilde X_n(\vect z) \big]^* = \lim_{t\to\infty} \nabla \widetilde \Phi_{F}(t, \vect z). \label{E:s6-4}
\end{align}
The integrals of $\widetilde X_i(\vect z)$ are given by
\begin{equation} \label{E:s6-3}
    \widetilde m_{i}(\vect z) = \lim_{t\to\infty}  (\widetilde \Phi_{F}(t, \vect z))_i - t,
\end{equation}
where the constant $-t$ is added before $t$ is taken to the limit, allowing the convergence of the limit.
Computing the exponential of the logarithms of Koopman eigenfunctions $\widetilde m_i(\vect z)$ and changing back to $\vect x$-coordinates, we obtain the formula for the principal Koopman eigenfunctions~\cite[Defn.~4]{kvalheim2021existence} on the region of attraction $A$ as stated below.

\begin{prop} \label{prop_formulas}
Assume Assumption~\ref{assum} holds.
The following expression converges uniformly to the principal Koopman eigenfunction with eigenvalue $\mu_i = \lambda_i$ for $i = 1,\ldots,\, n$ on every compact subset $K \subset A$, 
\begin{equation} \label{E:s6-9}
    \psi_i(\vect x) = \lim_{t \to \infty} e^{-\lambda_i t} \vect w_i^* \Phi_F(t, \vect x).
\end{equation}
\end{prop}

The proof is given in Appendix~\ref{proof_formulas}.

\begin{rem}
Eq. \eqref{E:s6-9} can be written in the form of variation of constants as
\begin{equation*}
    \psi_i(\vect x) = \vect w_i^* \vect x + \int_0^\infty e^{-\lambda_i s} \vect w_i^* F_n(\Phi_F(s, \vect x))\, ds
\end{equation*}
with $F_n(\vect x) = F(\vect x) - \nabla F(\vect x_0) \vect x$, which is the path-integral formula proposed by Deka et al. in~\cite{deka2023path}. In comparison, the advantage of (\ref{E:s6-9}) is replacing the limit of an integral by the limit of a function. On the condition for convergence, compared to~\cite{deka2023path} which proves convergence by assuming a small spectral gap, Assumption~\ref{assum} requires that the linearized system is non-resonant, which holds for a broader class of systems, e.g., systems with time-scale separation. The non-resonance assumption here is necessary for the existence of the Poincar\'e-Dulac normal form around an equilibrium point~\cite{arnold2012geometrical, kvalheim2021existence}, and a simple resonance example can be used to show that (\ref{E:s6-9}) generally fails to converge in the resonance case. However, it should also be noted that non-resonance is not necessary for the existence of smooth Koopman eigenfunctions on any open subset of the region of attraction bounded away from the equilibrium~\cite{kvalheim2025global}, where the flow-box theorem applies.
\end{rem}

\begin{rem}
Eq. (\ref{E:s6-9}) is a particular case of the computation formula in~\cite[Prop.~2]{mauroy2013isostables} for the isostable, which is formally defined through a Laplace average. An isostable is a level set of a principal Koopman eigenfunction such that all points on the isostable converge to the equilibrium at the same exponential rate.  While the isostable formula is well-known in practice, its convergence is usually not guaranteed. The symmetry-based characterization and the sufficient condition provided in Assumption~\ref{assum} are both steps toward analytical characterization of Koopman eigenfunctions and isostables.
\end{rem}

\begin{figure}[!t]
\centering
\includegraphics[scale=0.5,margin=0in 0in 0in 0.05in]{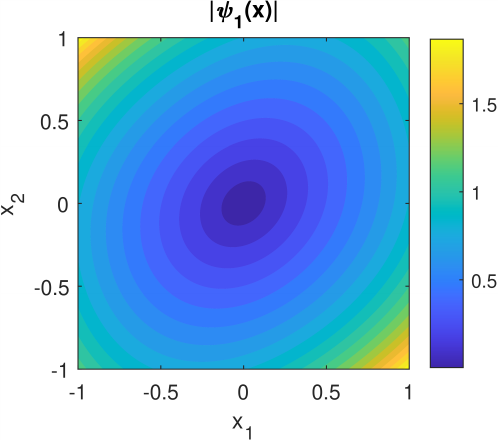}
\includegraphics[scale=0.5,margin=0in 0in 0in 0.05in]{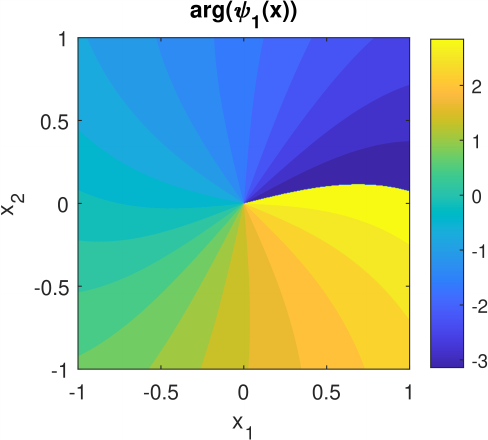}
\caption{Filled contours of the two principal Koopman eigenfunctions for the reverse-time van der Pol oscillator with $\mu = 0.5$, computed via (\ref{E:s6-9}). Trajectories are simulated with ode45 from a rectangular grid with $0.004$ spacing for $10$ s; this achieves $\leq 1\%$ maximum relative difference over the square domain compared with simulating for $8$ s.}
\label{fig_eigen}
\end{figure}

\begin{example}
To demonstrate the practicality of (\ref{E:s6-9}), we consider the example of the reverse-time van der Pol oscillator.
Using MATLAB, we verify the limiting expression (\ref{E:s6-9}) by computing the left eigenvectors and evaluating it over a grid of initial conditions. Both numerical convergence and the eigenfunction property are verified. The magnitude and phase of the two complex-conjugate principal eigenfunctions are plotted on the set $[-1, 1]^2$, as shown in Fig.~\ref{fig_eigen}.
\end{example}

\section{Conclusion} \label{sec_conclusion}
 
This paper establishes the equivalence between Koopman eigenfunctions and commuting symmetries as a special case of the equivalence between commuting and integrable local frames. This equivalence transforms the problem of identifying Koopman eigenfunctions—traditionally approached by fitting integrable linearizing vector fields, e.g., EDMD—into the problem of identifying commuting symmetry generators that form a local frame. We demonstrate the importance of this equivalence in analysis by deriving an explicit formula for the principal Koopman eigenfunctions on the region of attraction of a locally asymptotically stable equilibrium point. In doing so, we also demonstrate the correspondence between Koopman eigenfunctions and commuting symmetries on the region of attraction. 





\appendix

\subsection{Proof of Theorem~\ref{prop_equiv}} \label{proof_prop_equiv}

Let $\{\theta^1,\dots,\, \theta^n\}$ be the dual coframe to $\{E_1,\ldots,\, E_n\}$, so $\theta^i(E_j)=\delta^i_j$ and
\begin{equation*}
    \begin{bmatrix}\theta^1\\ \vdots\\ \theta^n\end{bmatrix} = \big[E_1\; \cdots\; E_n\big]^{-1}.
\end{equation*}
Recall Cartan’s identity for one-forms:
\begin{equation*}
    d\alpha(X,Y) = X(\alpha(Y)) - Y(\alpha(X)) - \alpha([X,Y]).
\end{equation*}

\emph{(Commuting $\Rightarrow$ conservative).}
If $[E_j,E_k]=0_n$, then for each $i$,
\begin{equation} \label{E:s6-22}
    d\theta^i(E_j,E_k) = -\theta^i([E_j,E_k]) = 0.
\end{equation}
Since $\{E_i\}$ is a frame, (\ref{E:s6-22}) implies $d\theta^i=0$. Since $U\subset \mathbb{R}^n$ is simply connected, there exists $z_i:U\to\mathbb{C}$ such that
\begin{equation*}
  \theta^i = dz_i .
\end{equation*}
Let $X_i=\nabla z_i$ so that $\theta^i(Y) = X_i^* (Y)$. Then
\begin{equation*}
    \big[X_1\; \cdots\; X_n\big]^* = \begin{bmatrix}dz_1\\ \vdots\\ dz_n\end{bmatrix}
    = \big[E_1\; \cdots\; E_n\big]^{-1},
\end{equation*}
so $\{X_i\}$ is a conservative frame.

\emph{(Conservative $\Rightarrow$ commuting).}
Conversely, if $X_i=\nabla z_i$ and $\theta^i = dz_i$, let $\{E_i\}$ be the dual frame to $\{\theta^i\}$, 
so $\big[X_1\; \cdots\; X_n\big]^*= \big[E_1\;\cdots\; E_n \big]^{-1}$. Since $d\theta^i=0$,
\begin{equation} \label{E:s6-17}
    0=d\theta^i(E_j,E_k) = -\theta^i([E_j,E_k]).
\end{equation}
Since $\{\theta^i\}$ is a coframe, (\ref{E:s6-17}) forces $[E_j,E_k]=0_n$ for all $j,k$.
\hfill \QEDclosed

\subsection{Proof of Proposition~\ref{prop_main1}} \label{proof_prop_main1}

\emph{(Linearizing $\Rightarrow$ symmetry).}
Write $X_i = \nabla z_i$ and $\theta^i = dz_i$ so that $\theta^i(Y) = X_i^*(Y)$. 
If $X_i^*F = \theta^i(F)=c_i$ (constant) and $d\theta^i=0$. 
Let $\{E_i\}$ be the dual frame to $\{\theta^i\}$. Then,
\begin{equation} \label{E:s6-20}
    0 = d\theta^i(F, E_j) = -\theta^i([F, E_j])
\end{equation}
Since $\{\theta^i\}$ is a coframe, (\ref{E:s6-20}) forces $[F,E_j]=0_n$ for all $j$. Commutativity $[E_j,E_k]=0_n$ already
follows from Theorem~\ref{prop_equiv} because $d\theta^i=0$. Hence, $\{E_i\}$ is a commuting symmetry frame.

\emph{(Symmetry $\Rightarrow$ linearizing).}
Conversely, suppose $\{E_i\}$ is a commuting symmetry frame: $[E_j,E_k]=0_n$ and $[F,E_j]=0_n$ for all $j,k$.
By Theorem~\ref{prop_equiv}, there exist closed one-forms $\theta^i$ with $\theta^i(E_j)=\delta^i_j$,
and (since $U \subset \mathbb{R}^n$ is simply connected) $\theta^i = dz_i$ with $X_i = \nabla z_i$ and $[X_1\;\cdots\; X_n]^* = [E_1\;\cdots\; E_n]^{-1}$. Then,
\begin{equation} \label{E:s6-21}
    E_j(\theta^i(F)) = F(\theta^i(E_j)) - d\theta^i(F, E_j) - \theta^i([F, E_j]) = 0.
\end{equation}
Since $\{E_j\}$ is a frame, (\ref{E:s6-21}) implies $\theta^i(F) = c_i$ (constant). 
Therefore,
\begin{equation*}
    \nabla z_i(x)^* F(x)=c_i,
\end{equation*}
which is exactly the linearizing condition.
\hfill \QEDclosed

\subsection{Proof of Proposition~\ref{prop_formulas}} \label{proof_formulas}


Assume, through a possible linear change of coordinates, that $\matr W = \matr I$.

By the non-resonance assumption from Assumption~\ref{assum}, for any $N\ge 2$ there exists a near-identity 
analytic conjugacy $T^{(N)}$ that eliminates all terms of degrees $2,\ldots,\, N$; see, e.g.,~\cite{arnold2012geometrical}. Hence, in the 
coordinates $\vect y := T^{(N)}(\vect x)$ the vector field takes the form 
\begin{equation*}
    G(\vect y) := \nabla T^{(N)}(\vect x)  F(\vect x) \big|_{\vect x = (T^{(N)})^{-1}(\vect y)} = \matr \Lambda \vect y + R^{(N+1)}(\vect y)
\end{equation*}
where
$R^{(N+1)}(\vect y) = \mathcal O(\|\vect y\|^{N+1})$.

Define $g_i^{(N)}(\vect y):= \vect w_i^*\, (T^{(N)})^{-1}(\vect y)$. Since $(T^{(N)})^{-1}$ is analytic and tangent to the identity, we have
\[
g_i^{(N)}(\vect y) = y_i + \mathcal O(\|\vect y\|^2).
\]
Moreover, since $T^{(N)}$ conjugates $F(\vect x)$ to $\matr \Lambda \vect y + R^{(N+1)}(\vect y)$ up to order $N$, we obtain the cohomological identity
\begin{equation} \label{E:id}
    \nabla g_i^{(N)}(\vect y)^*\, \matr \Lambda \vect y - \lambda_i g_i^{(N)}(\vect y) = \mathcal O(\|\vect y\|^{N+1}).
\end{equation}
Let $\vect y(t)$ solve $\dot{\vect y} = \matr \Lambda \vect  y + R^{(N+1)}(\vect y)$ with $\vect y(0) = \vect y'$. Then
\begin{align*}
    &\frac{d}{dt} \Big[e^{-\lambda_i t} g_i^{(N)}(\vect y(t))\Big] = \\
    &e^{-\lambda_i t}\Big[\nabla g_i^{(N)}(\vect y(t))^* \big[\matr \Lambda \vect y(t)+R^{(N+1)}(\vect y(t))\big] - \lambda_i g_i^{(N)}(\vect y(t))\Big].
\end{align*}
Combined with the identity (\ref{E:id}), this simplifies to
\begin{equation*} 
\frac{d}{dt}\Big[e^{-\lambda_i t} g_i^{(N)}(\vect y(t))\Big]
= e^{-\lambda_i t} \mathcal O(\|\vect y(t)\|^{\,N+1}).
\end{equation*}
Since $\|\vect y(t)\|\le C e^{\beta t}\|\vect y(0)\|$ for some $\beta \in (\delta, 0)$ with $\delta = \max_j \Re\lambda_j$ and constant $C > 0$ (by Gr\"onwall lemma), the RHS is bounded by
\begin{equation} \label{E:s6-16}
    C e^{((N+1)\beta - \Re\lambda_i)t} \|\vect y'\|^{N+1}.
\end{equation}
Note that we can choose
$N > \max\Big\{2,\, \big\lceil \max_i  \frac{\Re\lambda_i}{\beta} \big\rceil - 1 \Big\}$ to
guarantee $(N+1)\beta - \Re\lambda_i < 0$, so the bound in (\ref{E:s6-16}) is integrable on $[0,\infty)$. Therefore, $e^{-\lambda_i t} g_i^{(N)}(\vect y(t))$ converges uniformly for $\vect y'$ in any compact subset of the normal-form chart as $t\to\infty$. Equivalently,
\begin{equation*}
    \psi_i(\vect x) = \lim_{t\to\infty} e^{-\lambda_i t} \vect w_i^* \Phi_F(t,\vect x)
\end{equation*}
exists on the same compact sets.

By local asymptotic stability, there exists a compact $U_0 \subset A$ contained in the normal-form chart, and $t_K > 0$
such that $\Phi_F(t, \vect x)\in U_0$ for all $\vect x\in K$, $t\geq t_K$.
Writing $\tau = t - t_K$ and $\vect x' =\Phi_F(t_K, \vect x)\in U_0$, the uniform convergence on $U_0$
and the finite $|e^{-\lambda_i t_K}|$ give the uniform convergence on $K$.
Finally, to verify the eigenfunction property, it holds that
\begin{align*}
    \psi_i(\Phi_F(t,\vect x)) &= \lim_{s\to\infty} e^{-\lambda_i s}\vect w_i^* \Phi_F\big(s,\Phi_F(t,\vect x)\big)\\
    &= \lim_{s\to\infty} e^{-\lambda_i (s+t)}\vect w_i^* \Phi_F(s+t, \vect x) = e^{\lambda_i t} \psi_i(\vect x),
\end{align*}
for all $t\geq0$,
so $\psi_i(\vect x)$ is a Koopman eigenfunction with eigenvalue $\lambda_i$.
\hfill \QEDclosed

\section*{Acknowledgment}

The authors acknowledge the use of ChatGPT to explore and refine some steps of the proofs in the Appendix. All proofs were independently checked and validated by the authors, who take full responsibility for their correctness.




\bibliographystyle{IEEEtran}
\balance
\bibliography{references}

\end{document}